\documentclass[showpacs,preprint,amssymb]{revtex4}
\usepackage{graphicx}
\usepackage{dcolumn}
\usepackage{bm}
\def\be{\begin{equation}} 
\def\ee{\end{equation}}
\def\bea{\begin{eqnarray}} 
\def\eea{\end{eqnarray}}
\def\line{\hbox to \hsize}    
\def\frac #1#2{{#1\over #2}}

\def\Psid{\Psi^{\dagger}}

\def \r{{\bf r}}

\def\bz{{\bar z}}

\def \ket #1{{\vert #1\rangle}}

\def \brak #1#2{{\langle#1\vert#2\rangle}}

\def\1{\mbox{\bf 1}}
\def\bm#1{\mbox{\boldmath$#1$}}

\begin{document}

\title{Fullerenes, Zero-modes, and Self-adjoint Extensions}

\author{ABHISHEK ROY}

\affiliation{University of Illinois, Department of Physics\\ 1110 W. Green St.\\
Urbana, IL 61801 USA\\E-mail: aroy2@uiuc.edu}

 \author{ MICHAEL STONE}

\affiliation{University of Illinois, Department of Physics\\ 1110 W. Green St.\\
Urbana, IL 61801 USA\\E-mail: m-stone5@uiuc.edu}

\begin{abstract}  

We consider the low-energy electronic properties of graphene cones in the presence of a global Fries-Kekul{\'e} Peierls distortion. Such cones  occur  in fullerenes as the geometric  response to the disclination associated with pentagon rings. It is well known that the long-range effect of the  disclination deficit-angle can be modelled in the continuum Dirac-equation  approximation by  a spin connection and a  non-abelian gauge field. We show here that to understand the bound states localized in the vicinity of a pair of pentagons  one must, in addition to the long-range topological effects of the curvature  and gauge flux,  consider the effect  the short-range  lattice disruption near the defect. In particular, the radial Dirac equation for   the lowest angular-momentum channel sees the defect as a singular endpoint at the origin,  and  the resulting operator possesses   deficiency indices $(2,2)$. The radial equation therefore   admits a four-parameter set of self-adjoint boundary conditions.   The values of the four parameters  depend on how the pentagons are distributed and  determine whether or not there are  zero modes or other bound states.

\end{abstract}


\maketitle

\section{Introduction}

The essential features of the valance and conduction bands of  planar  graphene \cite{geim} can be modelled as  a pair of 2+1 dimensional Dirac  fermions, one for each of the conical band touchings  that occur  at the points ${\bf k}={\bf K}$ and ${\bf k}={\bf K}'$ in the Brillouin zone \cite{wallace}.  A Fries \cite{fries} (or equivalently Clar \cite{clar})   Kekul{\'e}-structure distortion of the hexagonal graphene  lattice will couple the two 
Dirac points and introduce a mass gap.  If the phase of the ${\bf K}$-${\bf K}'$ coupling  can vary  with position, there may  exist   topologically stable vortex textures  that  are  associated with zero modes and charge fractionalization \cite{chamon,chamon1,ghaemi}.  


A  complex-valued  ${\bf K}$-${\bf K}'$ coupling  is   unlikely to occur spontaneously as a result of a simple Peierls distortion -- although it might   be artificially  engineered through proximity-effect coupling to vortices in a superconducting substrate \cite{ghaemi,ghaemi2}. 
An alternative and naturally occurring  form of vortex  may  be  provided by  local curvature-inducing  geometric defects such as pentagons and heptagons \cite{lammart,osipov,pachos,pachos2}.  A spherical  fullerene such as C$_{60}$ possesses twelve pentagonal defects and, even  in the absence of Kekul{\'e}  distortion,  its energy spectrum contains  six delocalized low-energy levels. These near-zero-energy levels   have long been  understood as the lattice relicts  of six  exactly  zero-energy modes  of the   continuum   Dirac hamiltonian \cite{gonzalez1,gonzalez2}.  The continuum  zero-modes   are those predicted by the index theorem for 2+1 dimensional  Dirac fermions moving  in a fictitious monopole magnetic field that mimics  the effect of the defects.  One   quarter of a Dirac unit of monopole flux  threads through each of the twelve pentagons,  and for C$_{60}$ this discretely-lumped field is sufficiently spread out that the resulting  energy spectrum is well approximated by a spatially uniform field \cite{gonzalez1,gonzalez2}.   There  are three units of flux altogether, and so the index theorem predicts three zero modes apiece for  the two Dirac fermion species  of planar graphene.

The fictitious flux through the closed surface of the fullerene molecule means that when the discrete lattice hamiltonian is approximated by  two  continuous-space   Dirac hamiltonians, the continuum wavefunctions  must be regarded as sections of a twisted  line bundle, the twist being  characterized by a  Chern number of $\pm 3$. When a purely real Kekul{\'e}  field is introduced on the fullerene, the threefold twist,  when coupled with a natural choice of  how we  introduce  the gauge holonomy,  allows  the real  ${\bf K}$-${\bf K}'$ coupling  matrix elements to be   perceived by the continuum approximation as being associated with a charge-$2$ complex Higgs field  possessing  a net winding number of six. 
Each pair of pentagons is then effectively a single  vortex, and each such  vortex should, by the Jackiw-Rossi-Weinberg  index theorem \cite{jackiw_81,weinberg}, contain at least one zero mode. The six low-lying ``zero modes'' should therefore survive the introduction 
of the Kekul{\'e} induced mass gap, and, in theory,  its principal effect should be to localize the previously extended states in the vicinity of the pentagonal defects. 

In practice, numerical investigation of the lattice spectrum \cite{pachos}  shows that although the ``zero modes'' are not immediately  destroyed by the introduction of the Kekul{\'e}  Higgs field, their energy is affected. Indeed as the ratio $h$ of the double bond to single bond hopping   evolves from zero to large values the  ``zero-modes''  cross  the finite-size gap from the negative to the positive part of the spectrum.  What  was not obvious   from the plots in \cite{pachos} is that at the same time the zero-mode wavefunctions   evolve from being tightly localized around the defects  to being {\it antilocalized\/}. Thus  these modes, while still  topologically interesting, are not acting as expected from  the  continuum Jackiw-Rossi-Weinberg
index theorem. This theorem  predicts that the energy of the zero modes will be unaltered  by the introduction of the Higgs field, and that they will be localized for any sign of the mass.

In this paper we will explore what feature  of the continuum approximation to the  lattice hamiltonian is responsible for this   behaviour. We will focus on  isolated pentagons and  isolated pairs of pentagons.  These disinclination defects  roll the sheet of graphene into a cone. Away from the tip of the cone the geometry is locally flat   and a  continuum twisted Dirac hamiltonian approximation should be  reliable.  Near the tip,  the exact form of the lattice becomes important.  The lattice  effects can be accounted for, however, by imposing  a boundary condition on the continuum wavefunction at the tip of the cone. The result is  the introduction of  a four-parameter family of self-adjoint extensions to the Dirac operator. General values of the extension    parameters explicitly break  the symmetry  required for the index theorem, and  suitable choices of the parameters  reproduce the numerically observed  phenomena.

\section{Planar graphene and the Dirac equation}

 \begin{figure}
\includegraphics[width=2.5in]{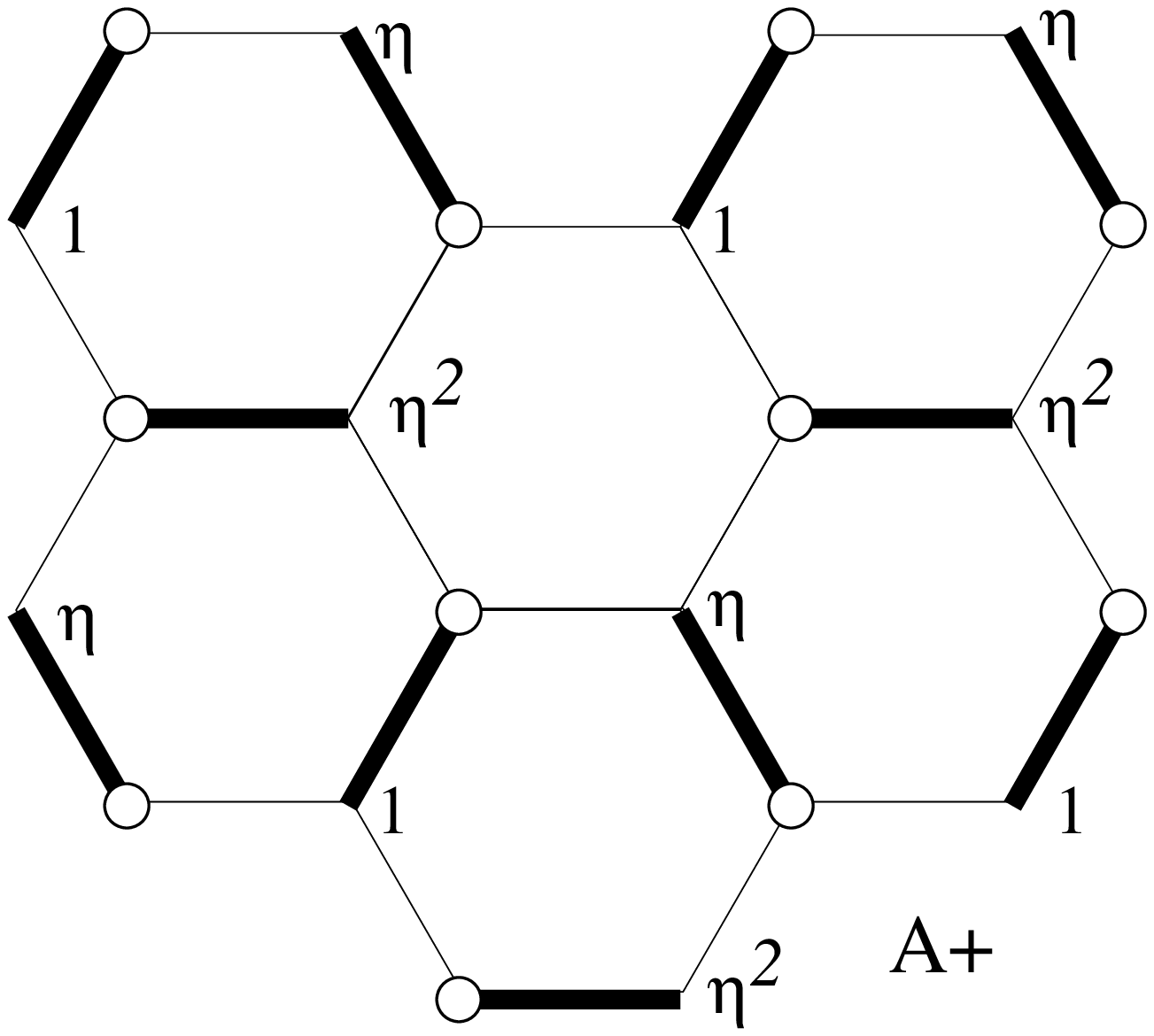}
\includegraphics[width=2.5in]{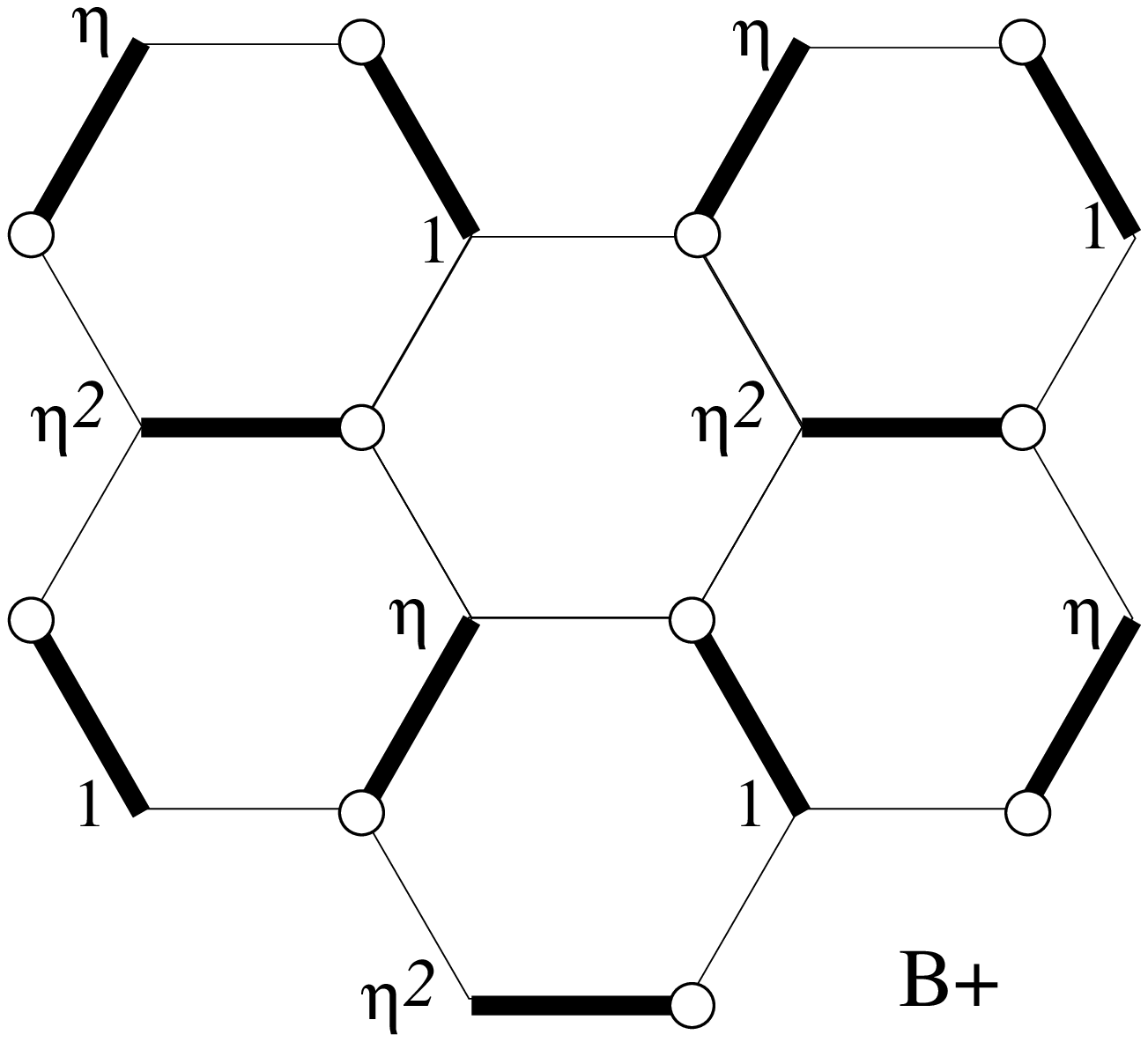}\\
\includegraphics[width=2.5in]{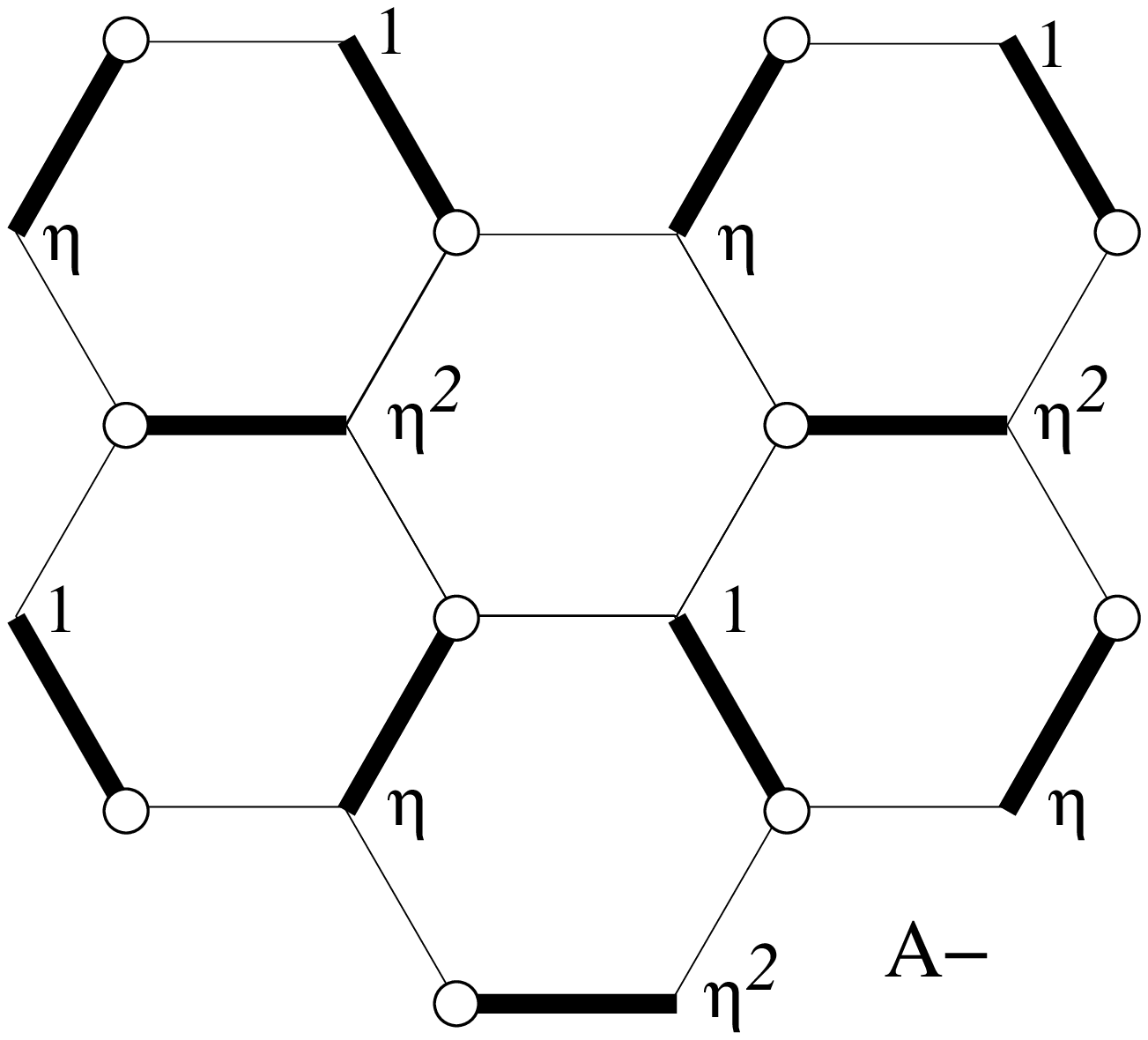}
\includegraphics[width=2.5in]{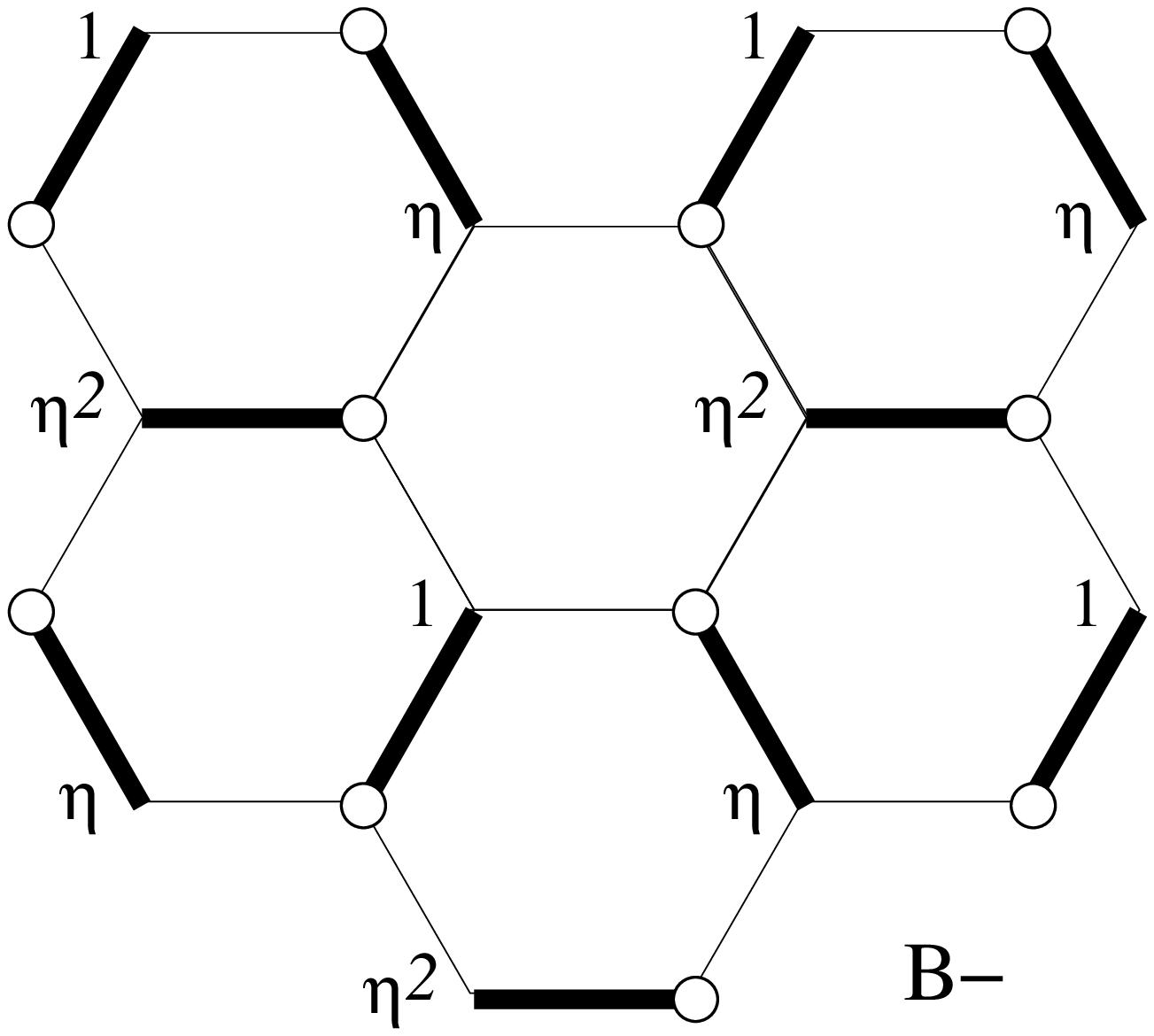}
\caption{The zero-energy reference states   $\ket{A+}$, $\ket{B+}$, $\ket{A-}$ and  $\ket{B-}$. The empty circles indicate that the wavefunction is zero at  those sites. The symbols $\eta$ and $\eta^2$  indicate that the wavefunction takes the values $\eta= \exp (2\pi i/3)$  and $\eta^2= \bar\eta= \exp (-2\pi i/3)$ at that site. The heavy lines indicate  a Fries stucture of  ``double'' bonds,  whose hopping will  increase  from $t$ to $t+\delta t$ when a Peierls  distortion occurs.}
\label{FIG:configs}
\end{figure}

In order to establish our notation, we begin with a quick  review of how the  tight-binding (H{\"u}ckel) approximation~\cite{gonzalez1} for planar graphene leads to  a massive Dirac hamiltonian.  In this approximation the $\pi$   electrons hop on a
two-dimensional honeycomb lattice with lattice constant $a$. The honeycomb lattice is bipartite with the property that  nearest-neighbour hopping  takes an electron  from sublattice A to sublattice B,  and vice versa.   We denote the  hopping  amplitude between nearest neighbour sites in undeformed graphene by    $t$. A  {\it Kekul{\'e} structure\/} is an  assignment  to the lattice edges of a pattern of single and double bonds  such that each   carbon atom  partakes of  two single bonds and one double bond. There are in general many such assignments,  but  patterns  that lead to a maximum number of benzene-like hexagons are known as {\it Fries structures\/}.  Fullerenes that possess a globally defect-free Fries structure (the so-called {\it leapfrog\/} fullerenes \cite{leapfrog1})  are typically the most  chemically stable.   This is because  a spontaneous Peierls distortion  that shortens the Fries-structure  double bonds and increases their hopping amplitude   from $t$ to $t+\delta t$  will  open a gap between the highest occupied molecular orbital (HOMO) and the lowest unoccupied  molecular orbital (LUMO),  lower the energy of the system, and give   the molecule the property of having a  ``closed shell.''  
Such a spontaneous distortion occurs in  C$_{60}$ where the double  bonds have length $\sim$ 0.1388 nm while  the single bonds have length $\sim $ 0.1432 nm \cite{rogers}.

The hopping hamiltonian for the undistorted infinite lattice has four linearly independent zero-energy eigenstates which are displayed  in figure \ref{FIG:configs}. The labels A and B indicate that the non-zero amplitudes are supported on sublattice A or B respectively, and the ``+'' and `$-$'' configurations have opposite handedness in the way that their  complex phases evolve as we circle the zero sites.  Both ``+''  configurations  have the same lattice momentum ${\bf  k}={\bf K}$ and both ``$-$''  configurations have lattice momentum ${\bf K}'$. Any low energy eigenfunction  will  be a slowly varying linear combination
\be
\Psi(\r) = f_{A+}(\r) \brak{\r}{A_+}+  f_{B+}(\r) \brak{\r}{B_+}+  f_{A-}(\r) \brak{\r}{A_-}+  f_{B-}(\r) \brak{\r}{B-}
\ee
of these reference  wavefunctions.
Here $\r$ labels discrete lattice points, but since the  $f(\r)$ are to be slowly varying on the scale of the lattice they can be regarded  as being smooth continuum  functions  ${\mathbb R}^2\to {\mathbb C}$.

The hamiltonian acts on the array of slowly varying functions $f$ as 
\be
H\left(\matrix{f_{A+}\cr f_{B+}\cr f_{B-}\cr f_{A-}}\right) =\frac{3at}{2}
\left(\matrix{0 &-2\partial_z & 0 &0\cr
                    2\partial_{\bz} &0 &0 & 0\cr
                    0 & 0 & 0 &2\partial_z\cr
                    0 & 0 & -2\partial_{\bz} &0}\right)\left(\matrix{f_{A+}\cr f_{B+}\cr f_{B-}\cr f_{A-}}\right) ,
\ee                   
where 
\be
\partial_z \equiv \frac{\partial}{\partial z}= \frac 12 \left(\frac{\partial}{\partial x}- i\frac{\partial}{\partial y}\right),\quad \partial_\bz \equiv  \frac{\partial}{\partial \bz}= \frac 12 \left(\frac{\partial}{\partial x}+i \frac{\partial}{\partial y}\right).
\ee
When the hopping is modified  by the Fries-Kekul{\'e} structure shown in figure \ref{FIG:configs}, the reference states are no longer exactly zero-energy    eigenfunctions. Instead $H\ket{A+} = \delta t \ket{B-}$ and $H\ket{A-}= \delta t \ket{B+}$.  Non-zero   matrix elements therefore appear coupling the  ${\bf K}$ Dirac point to the ${\bf K}'$ point:
\be
H(\Phi)\left(\matrix{f_{A+}\cr f_{B+}\cr f_{B-}\cr f_{A-}}\right) =
\frac{3at}{2} \left(\matrix{0 &-2\partial_z & \Phi &0\cr
                    2\partial_{\bz} &0 &0 &\Phi\cr
                    \Phi & 0 & 0 &2\partial_z\cr
                    0 & \Phi & -2\partial_{\bz} &0}\right)\left(\matrix{f_{A+}\cr f_{B+}\cr f_{B-}\cr f_{A-}}\right), 
\ee  
where $\Phi =(2/3at)\delta t$. The derivative blocks can be written as 
\be
\left(\matrix{ 0&- 2\partial_z\cr 2\partial_\bz &0}\right)= -i \sigma_2 \frac{\partial}{\partial x} +i\sigma_1  \frac{\partial}{\partial y}= \tilde {\bm \sigma} \cdot {\bf p},
\ee
where $\tilde {\bm \sigma}\equiv (\sigma_2,-\sigma_1)$ is a 90$^\circ$ rotated version of  the $ {\bm \sigma}= (\sigma_1,\sigma_2)$ spin vector.

We can if, we prefer, write
\be
H(\Phi)\left(\matrix{i f_{A+}\cr f_{B+}\cr if_{B-}\cr f_{A-}}\right) =
\frac{3at}{2} \left(\matrix{0 &-2i\partial_z & \Phi &0\cr
                    -2i\partial_{\bz} &0 &0 &\Phi\cr
                    \Phi & 0 & 0 & 2i\partial_z\cr
                    0 & \Phi & 2i\partial_{\bz} &0}\right)\left(\matrix{if_{A+}\cr f_{B+}\cr i f_{B-}\cr  f_{A-}}\right), 
\label{EQ:redefinesigma}                    
\ee  
in which case the derivative blocks involve  the more familiar form
\be
\left(\matrix{ 0&- 2i\partial_z\cr -2i\partial_\bz &0}\right)= -i \sigma_1 \frac{\partial}{\partial x} -i\sigma_2  \frac{\partial}{\partial y}= {\bm  \sigma}\cdot {\bf p}.
\ee
We thus    reduce the hamiltonian   to an unimportant constant times 
\be
H_{\rm Dirac} =  \left(\matrix{{\bm \sigma}\cdot {\bf p} & \Phi\cr \Phi&  -{\bm \sigma}\cdot {\bf p} }\right).
\ee
This  is a  two-dimensional reduction of the mass $m=|\Phi|$, three-dimensional Dirac hamiltonian. We notice that the matrix 
\be
\Gamma= \left(\matrix{\sigma_3  &0 \cr 0 &-\sigma_3}\right)
\ee
anti-commutes with $H_{\rm Dirac}$. This reflects the fact  that acting on a  wavefunction by $\Gamma$ inverts the sign of the wavefunction on one of the two sublattices  whilst  leaving the other sublattice alone, and that this involution  takes an  eigenfunction of energy $E$ to one of $-E$.  Zero energy states can be chosen so that they are eigenvectors  of $\Gamma$,  and the number of  $+1$  eigenvalues minus the number of $-1$  eigenvalues is related to topological data by the Jackiw-Rossi-Weinberg index theorem.

\section{Geometric defects and their gauge field}

\begin{figure}
\includegraphics[width=3.5in]{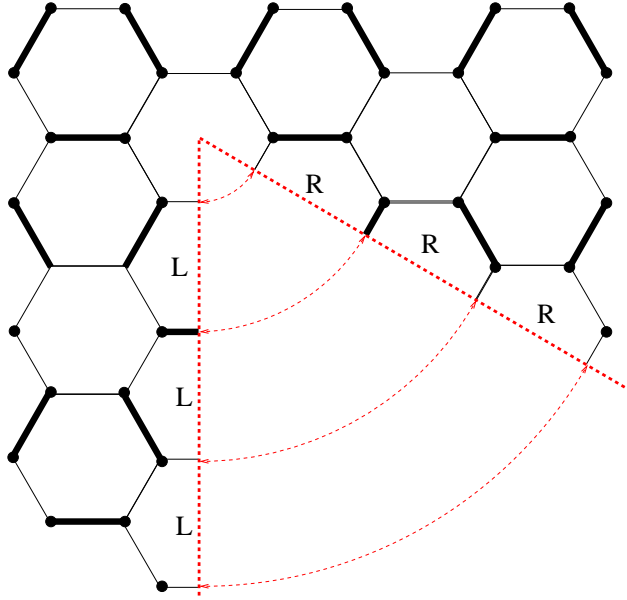}

\caption{Cutting out a $60^\circ$ wedge and reconnecting  the severed bonds leaves a pentagon disclination. When the apex of the wedge lies in a hexagon with no double bonds,  single bonds reconnect to single bonds and double bonds to double bonds, leaving a  globally consistent  Fries-Kekul{\'e}  structure.}
\label{FIG:wedge}
\end{figure}

We   insert a pentagon disclination into the lattice by cutting out a $60^\circ$ wedge and rejoining the severed bonds. We wish  to preserve the  global  Fries-Kekul{\'e}  pattern, so  the apex of the removed wedge must lie in  a hexagon with no double bonds (see figure \ref{FIG:wedge}).   In order for the lattice wavefunction to join smoothly across the seam of new bonds,  the coefficient  functions on the original lattice must obey the boundary conditions \cite{lammart,osipov}
\be
\left(\matrix{f_{A+}\cr f_{B+}\cr f_{B-}\cr f_{A-}}\right)_{\rm R} =
\left(\matrix{0&0&\bar \eta &0 \cr
0&0&0& \eta\cr
\bar\eta &0&0&0\cr
0&\eta& 0&0} \right)
\left(\matrix{f_{A+}\cr f_{B+}\cr f_{B-}\cr f_{A-}}\right)_{\rm L} .  
\ee
Here the subscript R means the value to   the right of the cut, and L to the left.  
In order to deal with this transformation it is convenient to change basis so that 
$
\Psi= (f_{A+} ,f_{B+},  f_{B-},  -f_{A-})^T.
$
This makes  $\pm$ kinetic energy blocks become identical
\be
H \left(\matrix{f_{A+}\cr f_{B+}\cr  f_{B-}\cr- f_{A-}}\right)=  \left(\matrix{0 &-2\partial_z & \Phi &0\cr
                    2\partial_{\bz} &0 &0 & -\Phi\cr
                    \Phi & 0 & 0 &-2\partial_z\cr
                    0 & -\Phi & +2\partial_{\bz} &0}\right)\left(\matrix{f_{A+}\cr f_{B+}\cr  f_{B-}\cr  - f_{A-}}\right)
\ee
at the expense of complicating the mass terms.  We can write the Hamiltonian matrix  compactly as
\be
H=({\mathbb I}\otimes \tilde {\bm \sigma})\cdot  {\bf p} +\Phi\, \tau_1\otimes  \sigma_3,
\ee
where the  $\mathbb I$ and   $\tau_1$ matrices  act on  the two-by-two blocks  (i.e. in   the  ${\bf K}$-${\bf K}'$ space) and the $\sigma$ matrices act between the $A$ and $B$ sublattices within each ${\bf K}$, ${\bf K}'$ block . (In the sequel we  will  omit   $\mathbb I$ matrices  when no ambiguity results.) 

In the new basis, the operator  $\Gamma$ that anti-commutes with $H$ remains 
$$
\Gamma =  \tau_3\otimes \sigma_3 \equiv    \left(\matrix{\sigma_3  &0 \cr 0 &-\sigma_3}\right). 
$$
The boundary condition, however,  becomes 
\be
\left(\matrix{f_{A+}\cr f_{B+}\cr  f_{B-}\cr -f_{A-}}\right)_{\rm R} =
\left(\matrix{0&0&\bar \eta &0 \cr
0&0&0& -\eta\cr
\bar\eta &0&0&0\cr
0&-\eta& 0&0} \right)
\left(\matrix{f_{A+}\cr f_{B+}\cr f_{B-}\cr-  f_{A-}}\right)_{\rm L}. 
\ee
This  can be factored as
\bea
\Psi_{\rm R} &=&-i\tau_1\otimes \exp\{-i\pi \sigma_3/6\} \Psi_{\rm L}\nonumber\\
&=& \exp\{-i\pi \tau_1/2\} \otimes \exp\{-i\pi \sigma_3/6\} \Psi_{\rm L}.
\eea
As before, the  $\tau_1$ matrix acts on the $\pm$  valley-degeneracy ``flavour''  indices  and the $\sigma$ matrices on the $A,B$ ``spin''  indices.  

Similarly, cutting out a 120$^\circ$ wedge turns a hexagon into a square,  and requires 
\be
\left(\matrix{f_{A+}\cr f_{B+}\cr  f_{B-}\cr -f_{A-}}\right)_{\rm R} =
\left(\matrix{\eta&0&0 &0 \cr
0&\bar\eta &0& 0\cr
0&0&\eta&0\cr
0&0& 0&\bar\eta} \right)
\left(\matrix{f_{A+}\cr f_{B+}\cr f_{B-}\cr-  f_{A-}}\right)_{\rm L} 
\ee
which can be factored as 
\bea
\Psi_{\rm R}&=&- {\mathbb I}\otimes  \exp\{-i\pi \sigma_3/3\} \Psi_{\rm L}\nonumber\\
&=& \exp\{-i\pi  \tau_1\}\otimes  \exp\{-i\pi \sigma_3/3\} \Psi_{\rm L}.
\eea
In this last line, the $\tau_1$ matrix in the exponent can be replaced by $\tau_2$, $\tau_3$, or even by $\mathbb I$,  without altering the lattice boundary condition.  The authors of  \cite{pachos}   elected to take this  matrix to be  $\tau_3$ as this choice leads to a  continuum hamiltonian for  which the symmetry  required by the  Jackiw-Rossi-Weinberg index theorem appears manifest.  

We can remove the discontinuity across the 60$^\circ$ wedge by writing 
\be
\Psi(r,\theta) = \exp\left\{i\frac{\pi}{2}  \tau_1 \left(\frac {3 \theta}{5 \pi}\right) \right\}  \otimes \exp\left\{i \frac{\pi}{6}  \sigma_3 \left(\frac {3\theta}{5\pi  }\right) \right\}\tilde \Psi(r,\theta),
\label{EQ:wedgebc}
\ee 
and across the 120$^\circ$ wedge by writing 
\be
\Psi(r,\theta) = \exp\left\{i  \pi \tau_1 \left(\frac {3\theta}{4 \pi }\right)  \right\}  \otimes \exp\left\{i \frac{\pi}{3} \sigma_3 \left(\frac {3 \theta}{4 \pi} \right)\right\}\tilde \Psi(r,\theta).
\label{EQ:wedgebc2}
\ee
Again, in the second case,   we may replace the $\tau_1$ matrix by $\tau_2$, $\tau_3$ or ${\mathbb I}$.
In all  cases, the new field $\tilde \Psi(r,\theta)$ is continuous across the reconnected seam.  In the first case the angle $\theta$ is restricted to  the range $-\pi /6<\theta< 3\pi/2$, with the limiting values representing the same point on the graphene cone,  and in the second $\pi/6<\theta<3\pi/2$.

We wish to write the eigenvalue problem in  polar coordinates whose origin is at the tip  of the cone.  To do this we  use the identity
\bea
H&=&\tilde {\bm \sigma}\cdot  {\bf p}+\Phi\, \tau_1 \otimes \sigma_3\nonumber\\
&=&-i \{\tilde \sigma_1 \cos\theta +\tilde \sigma_2 \sin\theta\}\frac{\partial}{\partial r} -i\{-\tilde \sigma_1 \sin\theta +\tilde \sigma_2 \cos\theta\}\frac 1 r\frac{\partial}{\partial \theta}+\Phi \,\tau_1\otimes  \sigma_3\nonumber\\
&=&e^{ -\frac i 2  \sigma_3\theta }\left(-i \tilde  \sigma_1 \frac{\partial}{\partial r} -i\tilde \sigma_2 \frac 1r \left( \frac{\partial}{\partial \theta}- \frac i 2 \sigma_3\right)+\Phi\, \tau_1 \otimes \sigma_3\right)e^{ \frac i 2 \sigma_3\theta }.
\eea
The coefficients of the derivatives $ \sigma_{r,\theta}\equiv \tilde \sigma_{1,2}$   are now matrix-valued constants, but they pay for their constancy  by being   attached to a  moving zweibein frame ${\bf e}_r$, ${\bf e}_\theta$. The   $-i\sigma_3/2$  spin connection  is therefore  required  to cancel the effect of taking the frame-rotation matrix $\exp\left\{ i  \sigma_3\theta /2\right\}$ through the $\theta$ derivative. 

For planar graphene   we would now define a new field $\chi(r, \theta)=e^{ \frac i 2 \sigma_3\theta }\Psi(r,\theta)$ and find that it is  {\it antiperiodic\/} under $\theta
\to \theta+ 2\pi$. Because of the deleted  wedge, however,  we have to define     
\be
\chi(r,\theta)= \exp\left\{  i \sigma_3\left(\frac 12 +\frac 1{10}\right)\theta \right\}\tilde \Psi(r,\theta),
\ee
which is  {antiperiodic} under $\theta\to \theta + 2\pi\,( 5/6) $.
Taking into account eq.\ (\ref{EQ:wedgebc}), the eigenvalue problem $H\Psi=E\Psi$  for the  60$^\circ$ wedge can now be written 
\be
\left(-i  \tilde \sigma_1 \frac{\partial}{\partial r} -i\tilde \sigma_2 \frac 1r \left( \frac{\partial}{\partial \theta}- \frac i 2 \sigma_3+\frac{ 3i}{10} \tau_1 \right)+\Phi\, \tau_1 \sigma_3\right)\chi(r,\theta)=E\chi(r,\theta).
\label{EQ:wedgetau1}
\ee
Here the $i \tau_1(3/10)$ gauge connection comes from taking the $\exp\left\{i \tau_1  (3 /10) \theta \right\} $ matrix  appearing in equation (\ref{EQ:wedgebc}) through the $\theta$ derivative. For the 120$^\circ$ wedge we must set,
\be
\chi(r,\theta)= \exp\left\{  i \sigma_3\left(\frac 12 +\frac 1{4}\right)\theta \right\}\tilde \Psi(r,\theta),
\ee
which is antiperiodic under $\theta \to \theta + 2\pi\,(2/3)$.   
The corresponding eignevalue problem becomes 
\be
\left(-i  \tilde \sigma_1 \frac{\partial}{\partial r} -i\tilde \sigma_2 \frac 1r \left( \frac{\partial}{\partial \theta}- \frac i 2 \sigma_3+\frac{ 3i}{4} \tau_1 \right)+\Phi\, \tau_1 \sigma_3\right)\chi(r,\theta)=E\chi(r,\theta).
\label{EQ:wedgetau2}
\ee

From now on, for purely cosmetic reasons, we will make the rotation introduced  in (\ref{EQ:redefinesigma}) that removes the tildes from the  $\sigma_{1,2}$  matrices. This does not affect the gauge fields, as they act in the valley degeneracy ``flavour''  space and the redefinition acts in the A-B ``spin'' space. 
 
We define   a new angle  $\phi = ( 6/5)\theta $,  or $\phi = (3/2) \theta$  that has the usual $2\pi$ periodicity.  ($\phi$  is the polar angle  seen when  looking along the axis of  the  60$^\circ$  or 120$^\circ$  cones   from above their apex.)  Using $\phi$ we can then separate the radial and angular part of the wavefunction as 
\be
\chi(r,\phi) = e^{ij\phi}\chi(r)
\ee
where
 $j$ takes half-integer values
\be
j= \ldots, -\frac 32,-\frac 12,+\frac 12,  +\frac 32\ldots.
\ee  

Note that the hermiticity of the {\it radial\/}  part of the differential operator  with respect to the inner product 
\be
\brak{\Psi_1}{\Psi_2} \stackrel{\rm def}{=}\int_0^{2\pi}\!\!\!\!  \int_0^\infty \Psid_1 \Psi_2\,rdrd\phi
\ee
requires  a contribution from the  spin-connection term $-i\sigma_3/2$ that occurs in the  {\it angular\/} covariant derivative part: 
\be
-i\sigma_2 \frac 1 r \left(\frac{\partial}{\partial \theta}   -\frac i{2}\sigma_3\right) = -i\sigma_2 \frac 1 r\frac{\partial}{\partial \theta}  -\frac i{2r} \sigma_1.
\ee
The  $i\sigma_1/2r$ ensures that  
\bea
\left( -i\sigma_1 \frac{\partial}{\partial r} -\frac i 2 \sigma_1\right)^\dagger &=&    -i\sigma_1 \frac 1r \frac{\partial}{\partial r}r  +\frac i 2 \sigma_1\nonumber\\
&=&  -i\sigma_1 \frac{\partial}{\partial r} -\frac i 2 \sigma_1.
\eea

 Because $\tau_1$ commutes with $H$, we    can find solutions to (\ref{EQ:wedgetau1}) and (\ref{EQ:wedgetau2})  (without the tilde's) of the form
\be
\chi_+(r,\phi) = \left( \matrix {u(r)\cr v(r)\cr u(r)\cr v(r)}\right) e^{ij\phi}, \quad \chi_-(r,\phi) = \left( \matrix {u(r)\cr v(v)\cr -u(r)\cr -v(r)}\right)e^{ij\phi},
\ee
which are eigenvectors of $\tau_1$ with eigenvalue $\pm 1$. 
The functions $u(r), v(r)$ then  satisfy 
\bea 
-i\left(\frac {d}{dr} +\frac 1{2r} +\frac 1r \frac{(j\pm n/4)}{1-n/6)}\right) v  +\tau \Phi u &=&Eu,\nonumber\\
-i\left(\frac {d}{dr} +\frac 1{2r} -\frac 1r \frac{(j\pm n/4)}{1-n/6)}\right )u -\tau \Phi v &=& Ev.
\label{EQ:uveigenvalue}
\eea
Here $n=1$ for the 60$^\circ$  cone and $n=2$ for the 120$^\circ$ cone. The number $\tau=\pm 1$  denotes  the  eigenvalue of $\tau_1$.  Set
\be
\nu=  \frac{j+\tau n/4}{1-n/6}.
\ee
We have   continuous-spectrum eigenfunctions with $u,v$ of the form
\bea
\left(\matrix{u(r)\cr v(r)}\right)&=&\left(\matrix{ (\epsilon +\tau \Phi)J_{\nu-1/2}(kr)\cr ik J_{\nu+ 1/2}(kr)}\right), \quad  E= \epsilon \equiv  +\sqrt{k^2+\Phi^2},\nonumber\\
\left(\matrix{u(r)\cr v(r)}\right)&=&\left(\matrix{ik J_{\nu-1/2}(kr)\cr (\epsilon+\tau \Phi) J_{\nu+ 1/2}(kr)}\right) , \quad  E= -\epsilon, 
\eea
and also
\bea
\left(\matrix{u(r)\cr v(r)}\right)&=& \left(\matrix{ (\epsilon +\tau \Phi)J_{-(\nu-1/2)}(kr)\cr -ik J_{-(\nu+ 1/2)}(kr)}\right), \quad  E= \epsilon \equiv  +\sqrt{k^2+\Phi^2},\nonumber\\
\left(\matrix{u(r)\cr v(r)}\right)&=&\left(\matrix{-ik J_{-(\nu-1/2)}(kr)\cr (\epsilon+\tau \Phi) J_{-(\nu+ 1/2)}(kr)}\right) , \quad  E= -\epsilon.
\eea
The  first set of solutions is finite at the origin when $j> 1/2$ and the second is finite at the origin when $j< -1/2$.  For $j=1/2$ and  $\tau$ negative, the upper component of the first set of  solutions diverges at the origin, but no faster than $r^{-1/2}$, so it is   normalizable there.  Similarly the second set is locally normalizable for $j=-1/2$.  

For most values of $j$ and $\tau$, demanding nomalizability is sufficient to select the physically allowed solutions.
When $n=2$, however, and for  $j=1/2$, $\tau=-1$, we have $\nu=0$. We also have $\nu=0$ when $j=-1/2$, $\tau=+1$. In these cases, the eigenvalue equation becomes 
 \bea
 -i\left(\frac{d}{dr}+\frac 1{2r}\right)v + \Phi \tau u&=&Eu\nonumber\\
-i\left(\frac{d}{dr}+\frac 1{2r}\right)u - \Phi \tau  v&=&Ev.
\label{EQ:special}
\eea  
 The scattering solutions with $E=\pm\sqrt{k^2+\Phi^2}$ contain the Bessel function  $J_{1/2}(kr)=\sqrt{2\pi/kr} \sin kr $ and 
$J_{-1/2}(kr)=\sqrt{2\pi/kr} \cos kr  $,  both of which are  normalizable near the singular point at the origin. The equation is therefore in Weyl's   limit-circle class there, and  some additional boundary condition must be imposed to  select a complete, linearly independent, set of solutions \cite{yamagishi}.

 One reason for    the $n=2$, 120$^\circ$ wedge differing  from the $n=1$, 60$^\circ$ wedge is that  that the cutting and sewing of the graphene sheet in the former case preserves the A-B bipartite structure  of the lattice. There should therefore be a some operator that anti-commutes with the Hamiltonian. This property  is not  easy to see  in eq (\ref{EQ:wedgetau2}), but  becomes clearer if we elect to replace the gauge field term with $\tau_3$. Then  equation (\ref{EQ:wedgebc2}) defining the single-valued  field  $\tilde \Psi(r,\theta)$ becomes 
\be
\Psi(r,\theta) = \exp\left\{i  \pi \tau_3 \left(\frac {3\theta}{4 \pi }\right)  \right\}  \otimes \exp\left\{i \frac{\pi}{3} \sigma_3 \left(\frac {3 \theta}{4 \pi} \right)\right\}\tilde \Psi(r,\theta),
\label{EQ:wedgebc3}
\ee
and the eigenvalue  equation (\ref{EQ:wedgetau2})  is replaced by 
\be
\left(-i  \sigma_1\left( \frac{\partial}{\partial r} +\frac 1{2r}\right) -i\sigma_2 \frac 1r \left( \frac 32 \frac{\partial}{\partial \phi}+\frac{ 3i}{4} \tau_3 \right)+\Phi\,  \sigma_3(\tau_1\cos\phi +\tau_2\sin\phi)\right)\chi(r,\theta)=E\chi(r,\theta),
\label{EQ:wedgetau3}
\ee
with antiperiodic $\chi(r,\theta)$.
This is the  equation with a unit-winding-number vortex in the mass term that was considered in \cite{pachos}. It is formally of the form to which the Jackiw-Rossi-Weinberg index theorem applies: the  matrix-valued differential operator  manifestly anti-commutes with $\Gamma= \tau_3\otimes \sigma_3$, and the   index is the Hilbert-space trace of $\Gamma$.

 Eq (\ref{EQ:wedgetau3}) posseses    zero energy  solutions of the form
\be
\chi(r,\phi) = \left(\matrix{ e^{-i\phi/2}  u(r) \cr0\cr0\cr  e^{i\phi/2}v(r)  }\right), \quad \chi(r,\phi) = \left(\matrix{ 0\cr e^{-i\phi/2} v(r)  \cr e^{i\phi/2}u(r) \cr 0}\right),
\ee
which are eigenvectors of $\Gamma$ with eigenvalues $+1$ and $-1$ respectively. 
Ignoring any constraints imposed by boundary conditions they are   
\be
\Psi_{0+-}=\left(\matrix{ e^{-i\phi/2}  \cr0\cr0\cr i e^{i\phi/2} }\right)\frac 1{\sqrt{r}}e^{-\Phi r} , \quad  \Psi_{0--}=\left(\matrix{ 0\cr ie^{-i\phi/2}   \cr e^{i\phi/2} \cr 0}\right)\frac 1{\sqrt{r}}e^{-\Phi r},
\ee
and  
 \be
\Psi_{0++}=\left(\matrix{ ie^{-i\phi/2}  \cr0\cr0\cr  e^{i\phi/2} }\right)\frac 1{\sqrt{r}}e^{+\Phi r} , \quad  \Psi_{0-+}=\left(\matrix{ 0\cr e^{-i\phi/2}   \cr i e^{i\phi/2} \cr 0}\right)\frac 1{\sqrt{r}}e^{+\Phi r},
\ee
Only one pair will be normalizable, depending on the sign of $\Phi$.

Eq (\ref{EQ:wedgetau3}) also posses more general bound-state  solutions  for any value of $E$ in the range  $-|\Phi|< E<|\Phi|$. These are 
 \be
\Psi_{E,1}=\left(\matrix{ (E+\Phi)e^{i\phi/2} \cr i\kappa e^{i\phi_2}&\cr (E+\Phi)e^{-i\phi/2}  \cr i \kappa e^{-i\phi}}\right)\frac 1{\sqrt{r}}e^{-\kappa r} ,\quad \Psi_{E,2}=\left(\matrix{( E-\Phi)e^{i\phi}  \cr i\kappa e^{i\phi/2}\cr  (\Phi-E)e^{-i\phi/2} \cr -i \kappa e^{-i\phi/2}}\right)\frac 1{\sqrt{r}}e^{-\kappa r}
\ee
and
\be
\Psi_{E,3}=\left(\matrix{ (E+\Phi)e^{i\phi/2}  \cr- i\kappa e^{i\phi/2}\cr  (E+\Phi)e^{-i\phi/2} \cr- i \kappa e^{-\phi/2}}\right)\frac 1{\sqrt{r}}e^{+\kappa r} ,\quad \Psi_{E,4}=\left(\matrix{ (E-\Phi)e^{i\phi/2} \cr -i\kappa e^{i\phi/2}\cr  (\Phi-E)e^{-i\phi/2} \cr i \kappa e^{-i\phi/2}}\right)\frac 1{\sqrt{r}}e^{+\kappa r}.
\ee
Here  $\kappa= \sqrt{\Phi^2-E^2}$.    These solutions are not eigenfunctions of  $\Gamma$. Instead  
 $\Gamma$ acts  on them to give a solution with energy $-E$.  


Both the $E=0$ and  the more general  $-|\Phi|< E<|\Phi|$ solutions are square integrable in the neighbourhood of $r=0$. The same is true of  the scattering state solutions with $|E|>|\Phi|$. We need to impose some boundary condition at $r=0$ to select  from these solutions a complete  linearly-independent  set.  As a guide to the  form of this boundary condition we apply the Weyl-von-Neumann theory \cite{richtmyer}. The matrix-valued  differential operator determining the radial part of all these solutions is  
\be
H= -i\sigma_1 \left(\frac {d}{dr} +\frac 1 {2r}\right) +\Phi\, \tau_1\otimes \sigma_3, \quad r\in  [0,\infty).
\ee
If we restrict it  an initial domain that  contains only functions that vanish at $r=0$, the resulting linear  operator   has deficiency 
indices $(2,2)$.  The  restricted operator  therefore admits a four-parameter  family of 
self-adjoint  extensions. To determine  the  corresponding   boundary conditions, we evaluate $ 
\brak{\Psi}{HX}-\brak{H\Psi}{X}
$
on functions $\Psi=(\psi_1,\psi_2,\psi_3,\psi_4)^T$ and $X= (\chi_1,\chi_2,\chi_3,\chi_4)^T$ that are square integrable on $[0,\infty)$ with the measure $r\,dr$.  We find that
\be
\brak{\Psi}{HX}-\brak{H\Psi}{X}= [ -ir(\psi_1^* \chi_2 +\psi_2^* \chi_1 +\psi_3^*\chi_4+\psi_4^*\chi_3)]_0^\infty.
\ee
Since  $\Psi(r)$ and $X(r)$ are allowed to diverge as $r^{-1/2}$ near the origin but must tend to zero faster than $r^{-2}$ at infinity, the vanishing of the integrated-out part  requires that the  expression in parentheses tend to  zero at $r=0$. 
If  we impose  boundary conditions
\be
\left(\matrix{\psi_1\cr \psi_3}\right)=
 \left(\matrix{a & b\cr c&d}\right) \left(\matrix{\psi_2\cr \psi_4}\right), \quad r\to 0
\ee
on $\Psi(r)$,  the adjoint boundary conditions on $X(r)$ are determined by requiring that
\be
\lim_{r\to 0} [\psi_2^*(a^* \chi_2 +c^* \chi_4 +\chi_1) + \psi_4^*(b^* \chi_2 +\chi_3 +d^* \chi_4)]=0
 \ee
 for any $\psi_2$, $\psi_4$. Thus the adjoint boundary conditions are
\be
\left(\matrix{\chi_1\cr \chi_3}\right)=\left(\matrix{-a^* & -c^*\cr- b^*&-d^*}\right) \left(\matrix{\chi_2\cr \chi_4}\right),\quad r\to 0.
\ee
For $H$ to be self-adjoint, these boundary conditions must coincide with the boundary conditions imposed on $\psi$. We there  have that, as $r\to 0$, 
\be
\left(\matrix{\psi_1\cr \psi_3}\right)= \left(\matrix{iA & B+iC\cr -B+iC&i D}\right) \left(\matrix{\psi_2\cr \psi_4}\right).
\label{EQ:general_bc}
\ee
for real (possiby infinite) numbers $A$, $B$, $C$ and  $D$.  This relation  contains  the four real parameters required by the Weyl-von Neumann count,  and so is  the most general  possible self-adjoint boundary condition.

 \begin{figure}
\includegraphics[width=2.7in]{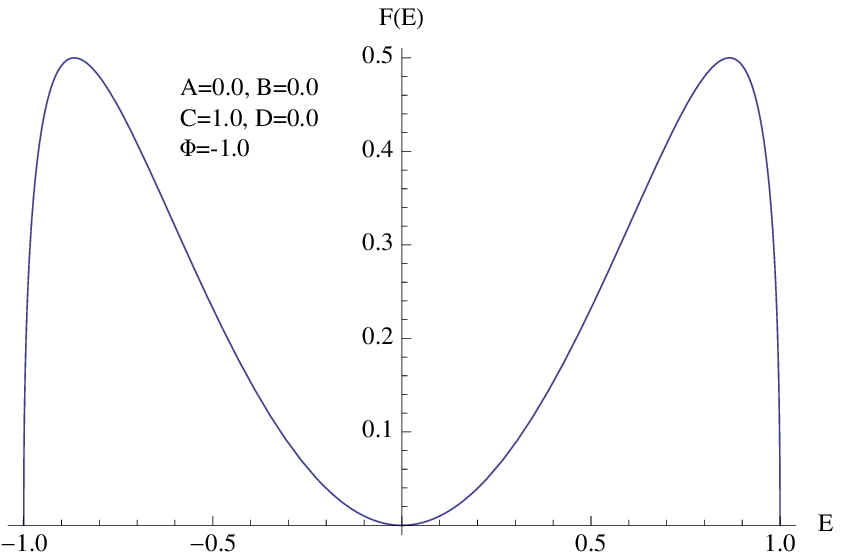}
\includegraphics[width=2.7in]{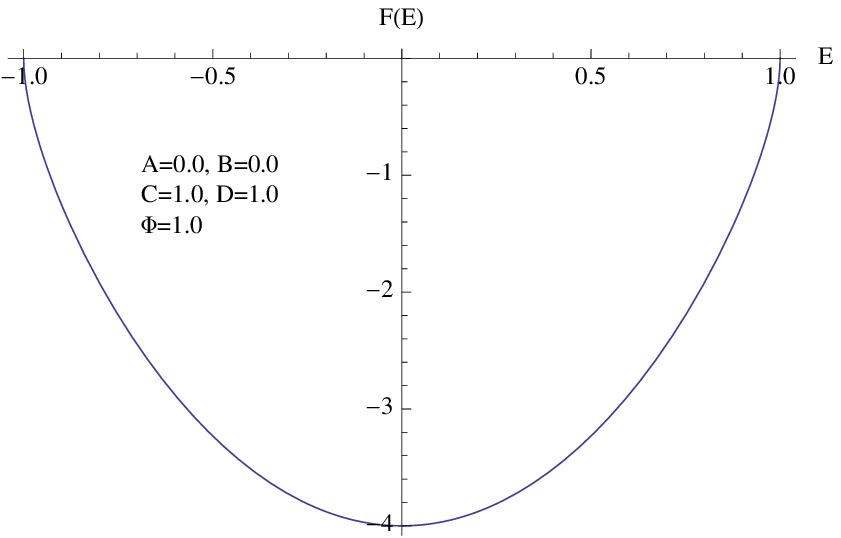}
\caption{The left-hand plot shows $F(E)$ for boundary-condition parameters   $A=B=D=0$, $C=1$,  and $\Phi=-1$ which possesses   two degenerate zero modes. 
The right-hand plot shows $F(E)$ for the same boundary parameters, but with  $\Phi$ having changed sign from negative (reduced hopping in the double bonds) to positive (enhanced hopping on the double bonds). There are now no bound states.}
\label{FIG:FE1and2}
\end{figure}

 \begin{figure}
\includegraphics[width=2.7in]{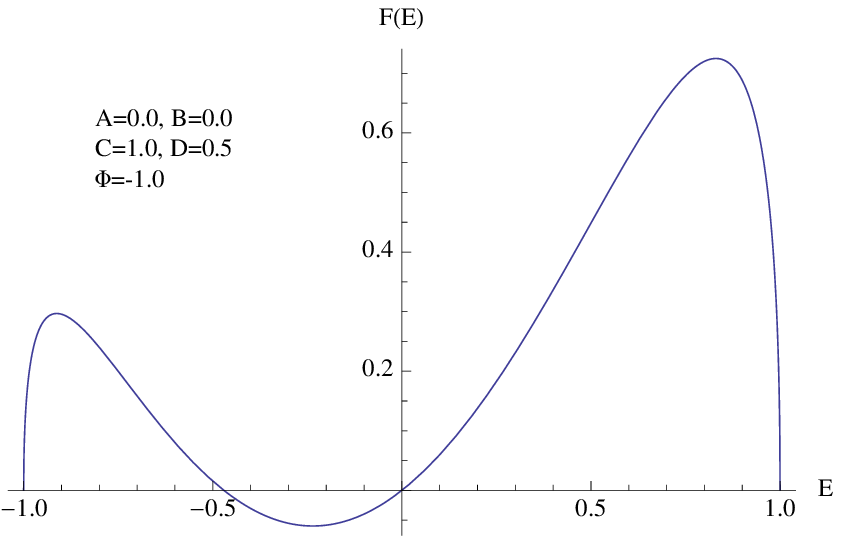}
\includegraphics[width=2.7in]{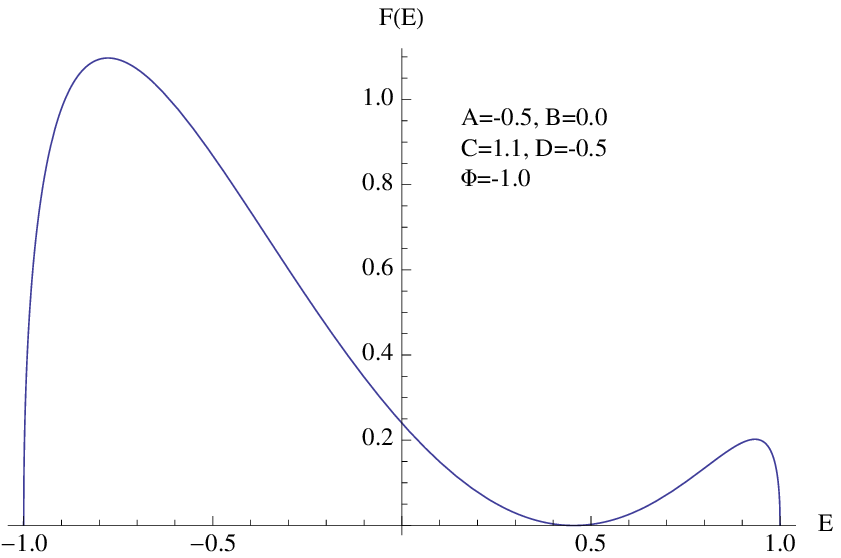}
\caption{The left-hand figure shows   $F(E)$ for $A=B=0$, $C=1$, $D=.5$ and $\Phi$ negative.  There is   a zero mode and a bound state with $E\approx -0.5 |\Phi|$. In the right-hand figure  we have $A=D=-1.0$, $B=0$, $D=\sqrt{2}$, and $\Phi$ negative. There are two  degenerate bound states at $E= |\Phi|/\sqrt{2}$. In both  cases,  the bound states cease to exist as soon as  $\Phi$ changes sign from negative to positive.}
\label{FIG:FE3and6}
\end{figure}

The numerical solution of the 120$^\circ$ wedge-cut lattice  show that there are two exact zero modes that are localized when $\Phi<0$ and become delocalized when $\Phi>0$.   These modes can be identified with $\Psi_{0++}$ and $\Psi_{0 -+}$, and  are allowed if we impose the boundary condition 
\be
\left(\matrix{\psi_1\cr \psi_3}\right)= \left(\matrix{0 & i\cr i & 0}\right) \left(\matrix{\psi_2\cr \psi_4}\right)
\label{EQ:zeroBCS}
\ee
at $r=0$.  The other potential zero modes $\Psi_{0+-}$ and $\Psi_{0 --}$ do not satisfy this  condition.

For general values of $A$, $B$, $C$ and $D$, we can seek bound-state solutions of the form 
$$
\Psi= \alpha \Psi_{E,1} +\beta \Psi_{E,2}.
$$
Imposing the boundary condition (\ref{EQ:general_bc}) at $r=0$ leads to a  pair of  homogeneous equations 
\bea
(\alpha+\beta) E+(\alpha-\beta)\Phi&=& -A\kappa (\alpha+\beta) -(C-iB)\kappa(\alpha-\beta),\nonumber\\
(\alpha-\beta) E+(\alpha+\beta)\Phi&=& -D\kappa (\alpha-\beta)-(C+iB) \kappa (\alpha+\beta) ,
\eea
for $\alpha$ and $\beta$, and hence to the condition $F(E)=0$, where
\be
F(E)=\left|\matrix{E+A\kappa & \Phi +(C-iB)\kappa\cr \Phi+(C+iB)\kappa& E+D\kappa}\right |.
\ee
The  function $F(E)$ is real in the range  $-|\Phi|\le E\le |\Phi|$, and always has two zeros at $E=\pm |\Phi|$ corresponding to the edges of the upper and lower continuum respectively. Additional zeros in the range $-|\Phi|<E<|\Phi|$ correspond to the energies of  bound states. Example plots of $F(E)$ for different values of $A$,  $C$, $D$ and $\Phi$ are shown in figures  \ref{FIG:FE1and2} and  \ref{FIG:FE3and6}.  Of particular interest  is  the  left hand plot  in figure  \ref{FIG:FE1and2}  which exhibits  the pair of    $E=0$ zero modes for the boundary conditions in (\ref{EQ:zeroBCS}), and the right-hand plot in figure \ref{FIG:FE3and6} which exhibits a  pair of degenerate levels  at a non-zero value of $E$.  The   plots in figure    \ref{FIG:FE3and6}  correspond to non-zero values for one or both of   the parameters $A$ and $D$. These parameters    control the   boundary coupling of the A and B lattices to  themselves, and  non-zero values  violate the bipartite lattice structure that leads to    the $E\leftrightarrow -E$ spectral symmetry possessed by the bulk differential equation.

 \begin{figure}
\includegraphics[width=4.0in]{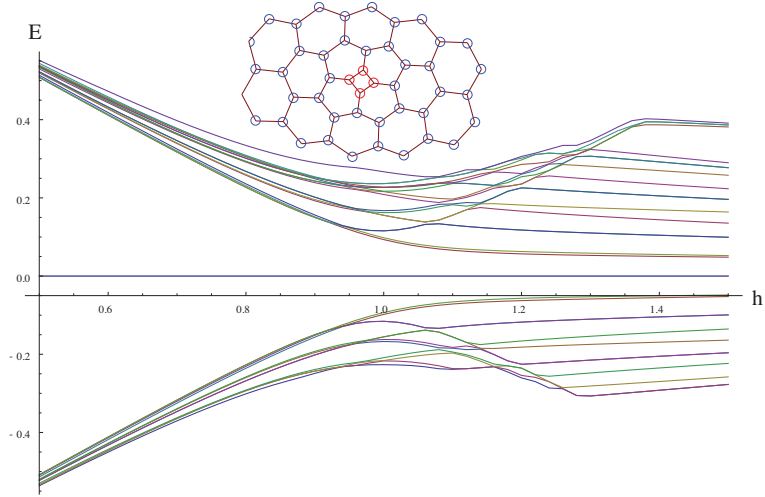}
\caption{ The low-lying part of the energy spectrum  plotted versus $h$ (the ratio of double bond hopping to single bond hopping) for a square defect created by excising a 120$^\circ$ wedge. The horizontal axis has been displaced vertically so as to uncover the doubly-degenerate exact zero mode.  The ``within-gap'' modes  peeling off from the upper and lower   continua for $h>1$ ({\it i.e\/}.\ $\Phi>0$) are ``anti-localized'' edge states,  whose exact form depends on how we truncate the lattice on its outer boundary.}
\label{FIG:squarespectrum}
\end{figure}

 \begin{figure}
\includegraphics[width=3.0in]{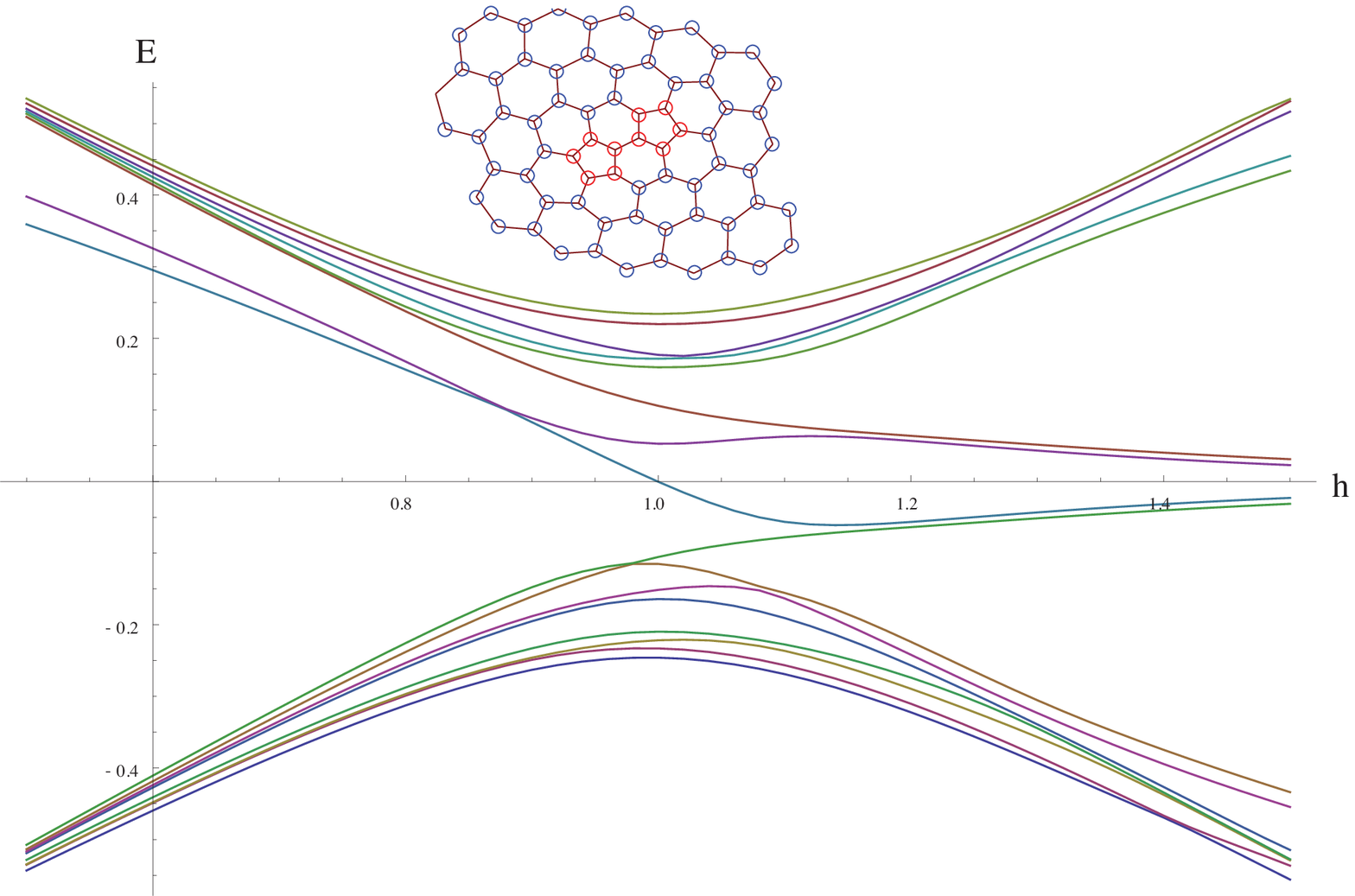}
\includegraphics[width=3.0in]{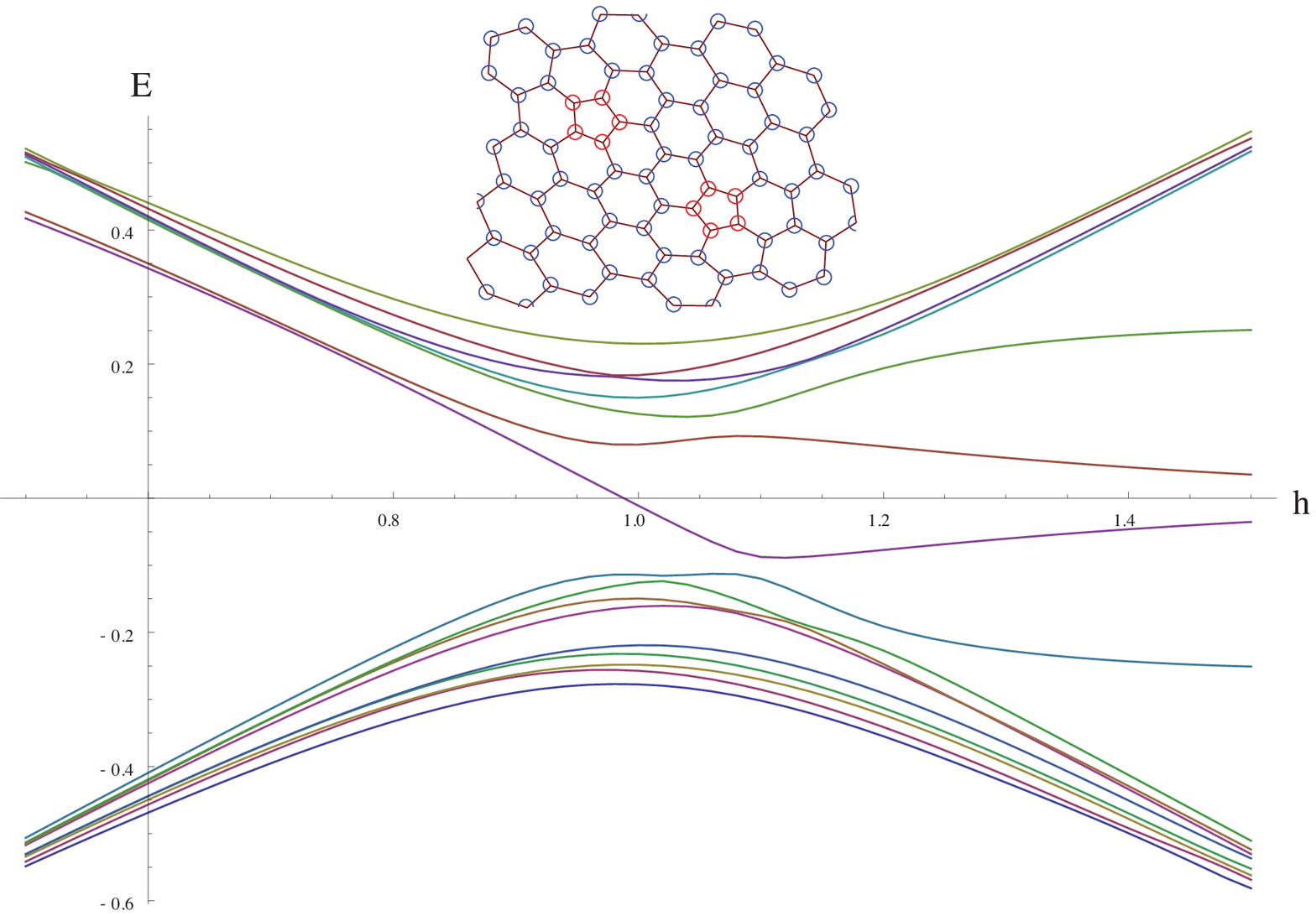}
\caption{The low-lying energy  spectrum  plotted versus $h$  for a pair of  nearby pentagons. The left-hand figure is for an $(n.m)=(1,1)$ cone in the languge of ref \cite{lammart}, and the right-hand figure is for an $(n,m)=(0,3)$ cone. In both cases the 60$^\circ$ wedges have their apices in  single-bond hexagons so as to preserve the global Kekul{\'e} structure. For  $h<1$ there is a pair of nearly degenerate bound states lying just below the upper continuum. The ``below gap"  modes  at $h>1$ are uninteresting edge states   localized at  the outer boundary.}
\label{FIG:pntspectrum1}
\end{figure}

 \begin{figure}
\includegraphics[width=4.5in]{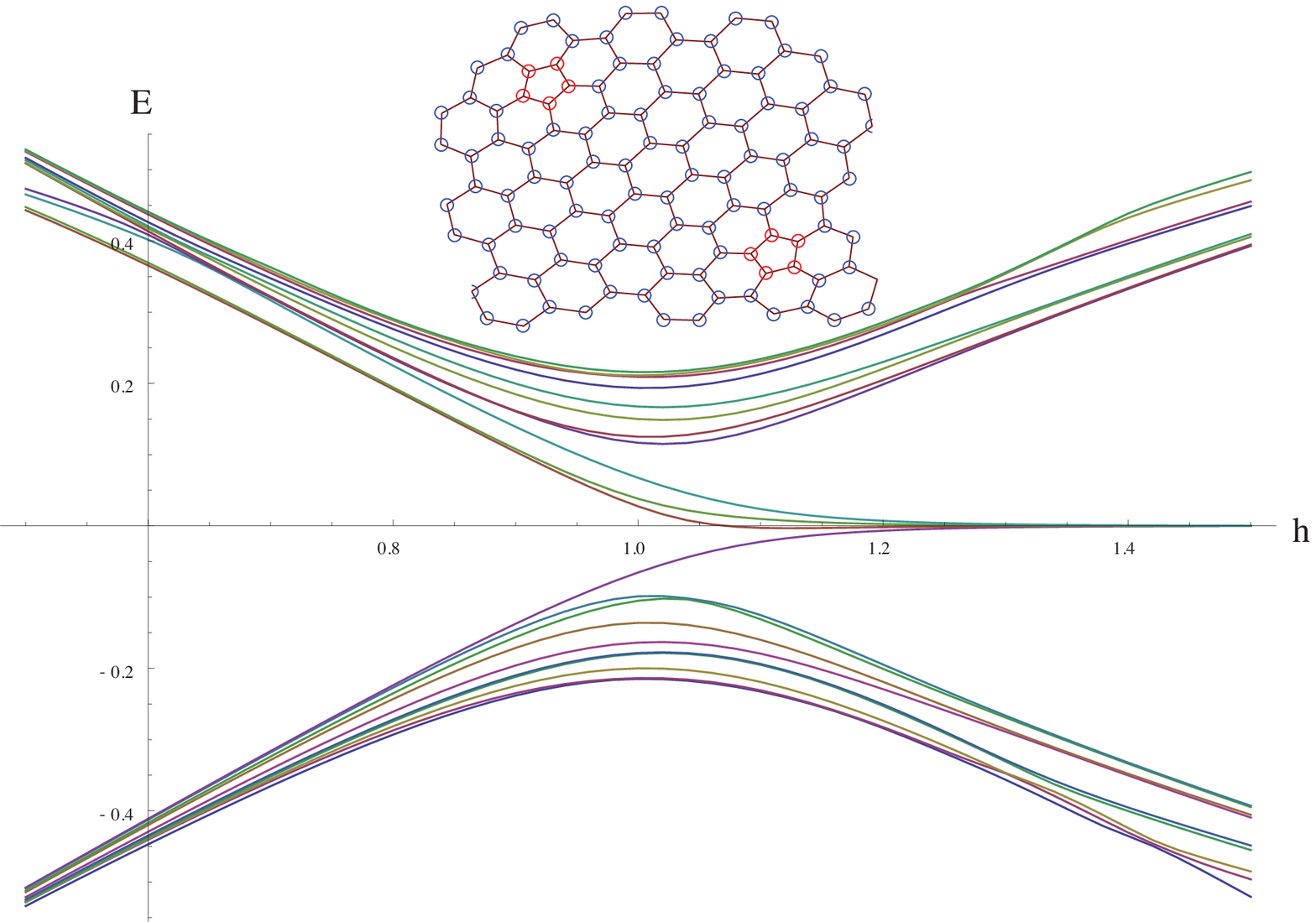}
\caption{The low-lying energy  spectrum  plotted versus $h$  for a pair of pentagons forming a  $(n,m)=(0,6)$ cone.}
\label{FIG:pntspectrum2}
\end{figure}

Figures \ref{FIG:squarespectrum},  \ref{FIG:pntspectrum1}  and  \ref{FIG:pntspectrum2} show some numerical plots of the low-lying energy states as a function of  $h= (t+\delta t)/t$, where $\delta t$ is the change in hopping parameter on the double bonds (recall that $\Phi=(2/3at)\delta t$.). 
In all four lattices,  the wedges have their apices in single-bond hexagons, so that the global Kekul{\'e} structure is preserved.   In the language of \cite{lammart} they all  have $n=m\, \hbox{\rm mod\,} 3$, ensuring that  the asymptotic  gauge field   is the same in all four  cases. The spectra differ, however, because the boundary conditions seen by the continuum wave functions at the tip of the cone are different.    In the case of figure  \ref{FIG:squarespectrum}, the global bipartite structure is preserved, the boundary conditions are those of  eq.\ (\ref{EQ:zeroBCS}) and so the spectrum is manifestly $E\leftrightarrow -E$ symmetric. In lattices of figures  \ref{FIG:pntspectrum1} and \ref{FIG:pntspectrum2} the bipartite structure is scrambled in the region between the pentagons and so the $E\leftrightarrow -E$ symmetry is violated. All of the two-pentagon  cases display  a pair of almost degenerate bound states just below the positive continuum,  and are  therefore similar to the spectrum associated with   the boundary conditions  of  the right-hand plot of   figure \ref{FIG:FE3and6}.  As  $h$ approaches unity, and $\Phi$ approaches zero, the wave functions begin to spread out and   see the outer  boundary of our large-but-finite lattices   (2644 vertices in case of the square and 1158 vertices for the pair of pentagons).  The effects of this can be seen in the figures 
in  the splitting of the bound-state energies as $h$ approaches unity from below.

\section{Conclusions}

The continuum Dirac  hamiltonian provides  a good account of the low-energy, long-wavelength, eigenstates of the tight-binding hamiltonian on an infinite sheet of graphene. The continuum model   is also a useful approximation for  cones tipped by curvature singularities induced by pentagon defects ---  but it must  be supplemented by non-trivial  boundary conditions at the tip of the cone. Although the low-lying   eigenfunctions have too long a wavelength to  resolve the fine details of lattice disruption      
at the tip, they  there experience phase shifts and mode mixing that have a significant effect on the eigenstates.   For two separated pentagons, the scrambling of the A-B bipartite lattice structure along a seam joining the  pentagons sufficiently violates the $E \leftrightarrow -E$ spectral symmetry as to allow bound states at non-zero $E$.  In the case of two coincident pentagons ({\it i.e.}\ a square) the  $E \leftrightarrow -E$ symmetry is preserved, but (in contrast to the quarter-unit flux through a pentagon)  the resultant half-unit of flux through the square plaquette is too large to be approximated by a  spread-out gauge field.  A continuum gauge field  with this flux would have bound a single state whose  eigenvalue of $\Gamma$ is determined by the sign of the flux. The lattice spectrum must be unchanged, however,  by the insertion of an integer flux-quantum through the square. Such an insertion can reverse the sign of the flux  and so no particular sign of $\Gamma$ can be favoured. The lattice hamiltonian compromises by producing two bound states, one with each sign of  $\Gamma$.  All these effects can be reproduced in the continuum model by suitable choices of the parameters in the self-adjoint boundary conditions.  However, the predictions \cite{pachos} of the continuum Jackiw-Rossi-Weinberg theorem do not  survive in a simple form.  It will be interesting to see if the theorem can be modified to include the effects of the singular  endpoint.

\section{Acknowledgements}

This work was supported by the National Science Foundadation  under grant DMR-06-03528. Some  of the work was carried at the KITP in Santa Barbara, and so  supported in part by the National Science Foundation under grant PHY05-51164. We would like to thank  Jiannis Pachos for many discussions, and  Jeremy Oon for his assistance  with the numerical work at the beginning of this project.

\end{document}